# Intensity-correlated spiking infrared and ultraviolet emission from sodium vapors

ALEXANDER M. AKULSHIN[1,2*], FELIPE PEDREROS BUSTOS[1,3], AND DMITRY BUDKER[1,4]

[1] *Helmholtz-Institut, GSI Helmholtzzentrum für Schwerionenforschung, 55128 Mainz, Germany*
[2] *Optical Sciences Centre, Swinburne University of Technology, Melbourne, Australia*
[3] *Aix Marseille Univ, CNRS, CNES, LAM, Marseille, France*
[4] *Department of Physics, University of California, Berkeley, CA 94720-7300, USA*
*Corresponding author: aakoulchine@swin.edu.au*

**The directional spiking infrared and ultraviolet emission from sodium vapors excited to the $4D_{5/2}$ level by a continuous-wave resonant laser pump, that constitute a novel feature of the cooperative effects, has been analyzed. Cascade mirrorless lasing at 2207 and 2338 nm on population-inverted transitions and ultraviolet radiation at 330 nm that is generated due to four-wave mixing process demonstrate a high degree of intensity correlation.**

http://dx.doi.org/10.1364/OL.99.099999

Under certain conditions, radiative dynamics of an ensemble of excited emitters synchronised through their spontaneous emission can radically differ from the usual radiative behaviour of an individual emitter due to the cooperative effects (CE), an extensive study of which was initiated by R. Dicke [1]. A synchronised population-inverted ensemble can relax to the ground state much faster than individual emitters radiating a short burst of directional emission. The effect of cooperative emission is also known as superfluorescence or superradiance, depending on initial conditions and peculiarities of excitation [2]. Thoroughly studied cooperative emission was reported in various systems ranging from phased antenna arrays [3] and Josephson junctions [4] to warm [5] and laser cooled atomic samples [6], active continuous media [7], quantum dots and diamond nanocrystals [8, 9].

The diversity of media of interest and methods of excitation explains the vast variety of manifestations of the collective effects, however, a delayed burst of directional emission with durations shorter than the natural lifetimes of the corresponding energy levels is a common feature of CE in any atomic medium in which a short pump pulse produces population inversion. In this letter, we demonstrate a novel feature of cooperative emission in alkali metal vapours with population-inverted transitions. Short bursts of directional emission can be generated even under continuous-wave (cw) laser pumping. In other words, short-pulse excitation is not a necessary condition for observing directional cooperative radiation.

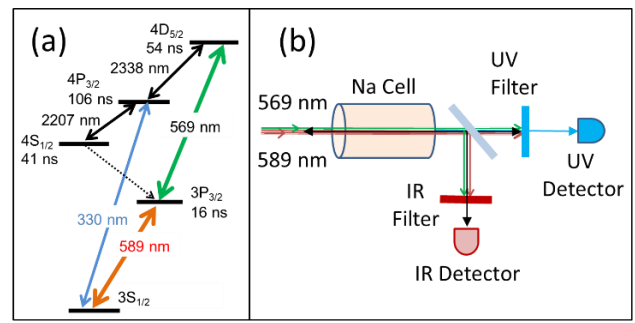

Fig. 1. a) Energy levels of Na atom involved in the combined stepwise and two-photon excitation by resonant laser light at 589 and 569 nm that results in new-field generation at 2338 nm, 2207 nm and 330 nm. b) Optical scheme of the experiment.

First observations of collective effects in alkali vapours were reported on IR transitions [5, 10]. In the visible domain, pulses of tens of picosecond in duration, significantly shorter than the time scale of the typical relaxation processes including spontaneous decay, were observed at 420 nm in Rb vapours excited by even shorter laser pulses [11-14]. The observed bursts of directional emission were attributed to superfluorescence. However, no evidence of population inversion at the corresponding transitions was provided, although this is an essential requirement for CE. In [15], where a spiking behaviour of directional emission at 420 nm in Rb vapours was reported, it was suggested that short pulses of the directional radiation at 420 nm are produced by a parametric wave-mixing process, in which the cw applied laser fields interact with cooperative emission not in the visible, but in the mid-infrared spectral region, generated on the population-inverted $6P_{3/2}$-$5D_{5/2}$ transition. However, in the experiment mentioned above, this radiation was not analysed with a high temporal resolution.

We note that despite frequency up- and down-conversion of resonant radiation in alkali vapours becoming an active topic of research [16-26], surprisingly little is known about temporal characteristics of the generated fields.

Here, we present the results of an experimental study of temporal properties of the optical fields generated by sodium atoms two-photon excited to the $4D_{5/2}$ level. Figure 1a shows the relevant energy levels of Na atoms involved in the interaction with applied laser fields at 589 and 569 nm, which are resonant to the $3S_{1/2}$-$4P_{3/2}$ and $3P_{3/2}$-$4D_{5/2}$ transitions, respectively. The main advantage of using Na atoms is that the generated fields are in the spectral regions where fast and sensitive detectors are readily available. Also, unlike Rb vapours, a special vapour cell with windows transparent for mid-IR radiation is not required. This makes possible not only a direct observation of the unexpected spiking regime of mirrorless lasing under cw pumping, but also enables correlation analysis of the generated optical fields.

The experimental setup for studying forward-directed emission of sodium vapour is practically the same as that described in our previous paper [22]. A simplified optical scheme of the experiment is shown in Fig. 1b. Monochromatic radiation at 589 and 569 nm used for Na atoms excitation is provided by two cw dye lasers, Coherent 699-21 and Coherent 899-21. Optical frequencies of the lasers are monitored with a wavemeter (High Finesse WS6-600) and a confocal Fabry-Perot (FP) cavity. The upper limit of the short-term laser linewidth of 5 MHz is estimated based on FP cavity transmission resonances recorded with the 10 ms sampling time. The intensity noise of laser light can be characterized by relative standard deviation (RSD) that is defined as RSD = $\sigma/\mu \times 100\%$, where $\sigma$ and $\mu$ are the standard deviation and the mean value, respectively. We find that both the pump laser fields have similar noise properties, RSD = (1.2 ± 0.2) %.

A 30-cm long sealed quartz cell containing metal sodium without buffer gas is used in the experiment. A resistive heater provides a temperature gradient along the cell preventing sodium atoms from condensing on the windows. The Na atom number density is estimated based on the measured temperature in the coldest part of the cell [27].

Sodium atoms are excited from the ground state to the $4D_{5/2}$ level with a co-propagating bichromatic beam that consists of laser radiation at 589 and 569 nm combined using a non-polarizing beam splitter. Before entering the cell, the bichromatic beam passes through a polarizer and a quarter-wave plate. The beam is focused inside the cell with a long-focus lens so that the minimal diameter of the 589 and 569 nm components are approximately 250 and 350 µm, respectively. The 589 nm pump laser is frequency modulated at 1.713 GHz with an electro-optical modulator, since velocity-selective ground-state hyperfine repumping enhances the excitation of atoms to the $4D_{5/2}$ level [28].

The generated IR and UV radiation transmitted through appropriate bandpass interference filters is detected with an InGaAs amplified photodiode PDA10D2 having a 25 MHz detection bandwidth and an alkali-cathode photomultiplier tube (PMT). Signals, which are proportional to the intensity of the new fields, are recorded with a digital oscilloscope having a 100 MHz detection bandwidth.

The generated forward-directed amplified spontaneous emission (ASE) at 2338 µm, as well as the previously studied ASE at 2207 nm, reveal pronounced threshold-type dependences on the atom number density $n$ in the cell and the pump-light power. For example, Fig. 2 shows that both IR fields rise sharply above a certain levels of applied pump radiation at 569 nm, while side-detected UV fluorescence is growing steadily over the entire range. It also demonstrates that the power threshold for mirrorless lasing at 2338 nm is even lower.

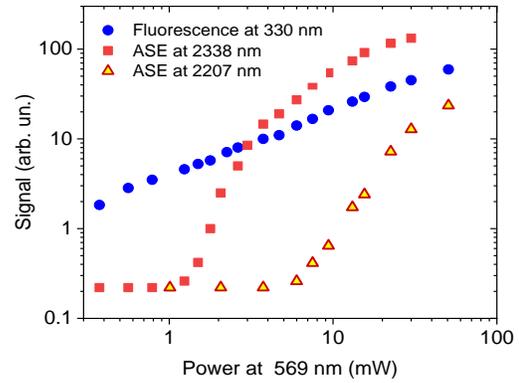

Fig. 2. Threshold-type dependences of cascade primary and secondary mirrorless lasing at 2338 nm and 2207 nm. Intensity of forward-directed emission at 2338 nm and 2207 nm, as well as UV fluorescence at 330 nm from sodium vapours excited to the $4D_{5/2}$ level with co-propagating laser beams as a function of the laser power at 569 nm, while the applied laser power at 589 nm is 18 mW. Data are taken with both lasers tuned to maximum mirrorless lasing intensity. The atom number density in the cell is $6\times10^{11}$ cm$^{-3}$.

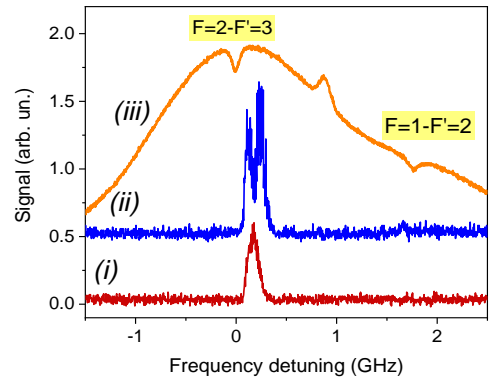

Fig. 3. Spectral dependences of (i) mirrorless lasing at 2338 nm and (ii) FWM at 330 nm as a function of frequency detuning of the pump laser at 589 nm from the $3S_{1/2}(F=2)$-$3P_{3/2}(F'=3)$ transition ($\Delta\nu_{589} = \nu_{589} - \nu_{23}$). Curve (iii), the spectral profile of saturated fluorescence at 589 nm in an axillary cell on the Na-D2 line is used as a frequency reference for the 589 nm laser. The fixed-frequency pump radiation at 569 nm is approximately 170 MHz blue-detuned from the $3P_{3/2}(F'=3)$-$5D_{5/2}$ transition. Curves (ii) and (iii) are shifted vertically for a better visibility.

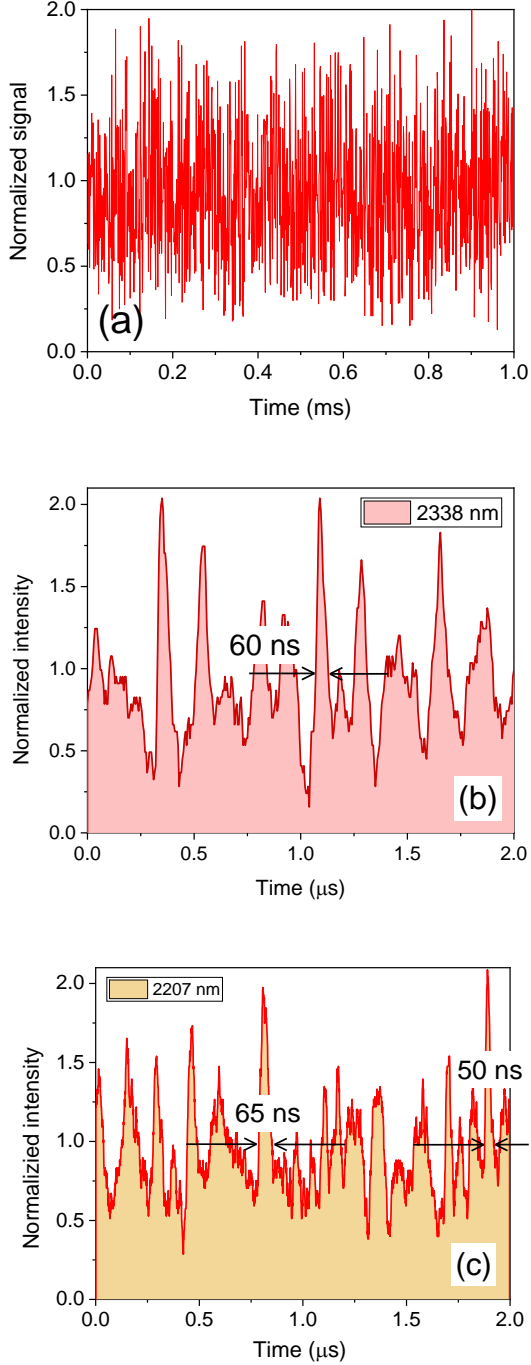

and FWM profiles demonstrate a sub-Doppler structure because of velocity-selective excitation to the $4D_{5/2}$ level as was discussed in [30]. Such process of parametric FWM was studied in detail in Rb and Cs vapours using similar excitation schemes [16, 18, 20].

Detecting the mirrorless lasing fields at 2338 and 2207 nm with a high temporal resolution, we find that they consist of chaotic, partially overlapping spikes despite the cw pump excitation to the $4D_{5/2}$ level, as shown in Fig. 4. The RSD of mirrorless lasing at 2338 nm shown in Fig. 4a is (40 ± 5) %, which is much higher than the intensity noise of the applied pump radiation. The temporal structure of intensity variations of both the generated IR fields are approximately the same. The time at which the spikes appear, as well as their shape, duration and amplitude, vary considerably from sample to sample. The shortest FWHM duration of the recorded spikes of the 2338 nm lasing is about 60 ns (Fig. 4b), which is close to the natural lifetime of the $4D_{5/2}$ level, whereas some spikes at 2207 nm recorded under the same experimental conditions have FWHM duration in the range of 50 - 65 ns, which is noticeably shorter than the natural lifetime of the $4P_{3/2}$ level (Fig. 4c).

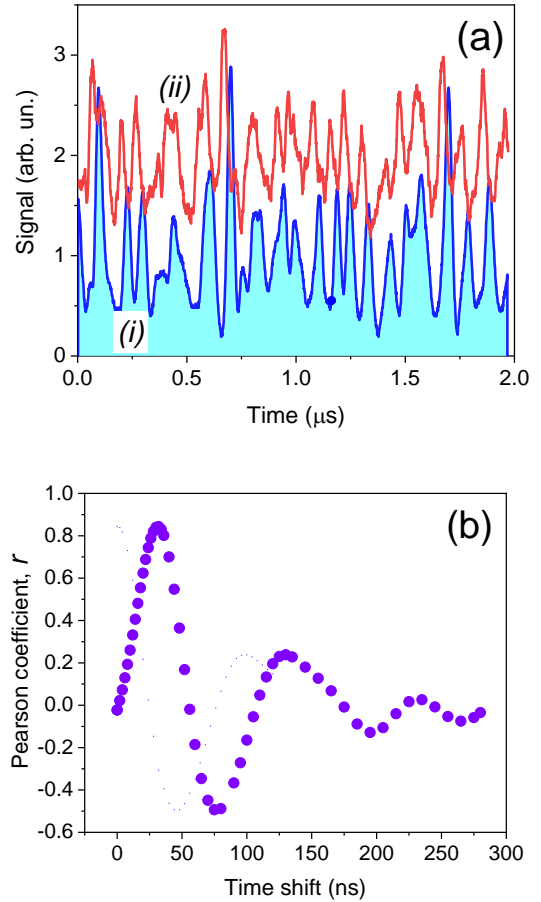

Fig. 4. Temporal intensity variations of mirrorless lasing at (a, b) 2338 nm and (c) 2207 nm, respectively, when the fixed-frequency lasers are tuned to the $3S_{1/2}(F=2)-3P_{3/2}(F'=3)$ and $3P_{3/2}(F'=3)-4D_{5/2}$ transitions. The applied cw pump power $P_{589}$ = 25 mW and $P_{569}$= 16 mW, while $n \approx 8.5 \times 10^{11}$ cm$^{-3}$.

We find that an interaction of the generated forward-directed IR radiation at 2338 μm with the applied co-propagating pump radiation at 589 and 569 nm in Na vapours produces a directional UV emission at 330 nm by the process of parametric four-wave mixing (FWM). Spectral dependences of the mirrorless IR lasing and the directional UV emission are shown in Fig. 3. Both the ASE

Fig. 5. (a) Temporal intensity profiles of (i) UV and (ii) IR emission at 330 nm and 2338 nm, respectively. The fixed-frequency pump lasers are tuned to the $3S_{1/2}(F=2) - 3P_{3/2}(F'=3)$ and $3P_{3/2}(F'=3) - 4D_{5/2}$ transitions, respectively, to maximize the generated UV intensity. Curve (ii) is shifted vertically on one unit for a better visibility. (b) Pearson coefficient as a function of the time shift between the profiles.

As has been mentioned above, a similar temporal behaviour in the case of cw excitation was observed with directional blue light at 420 nm generated in Rb vapours. These observations support the idea that spiking regime of the FWM field at 420 nm is due the temporal characteristics of the directional mid-IR emission generated on the population-inverted $5D_{5/2}$-$6P_{3/2}$ transition in Rb atoms [15].

We find that the directional UV emission also consist of partly overlapping spikes. This is natural, as spiking IR radiation is directly involved in preparing nonlinear polarization for the $3S_{1/2}(F=2)$-$3P_{3/2}$ transition, $P \sim \chi^{(3)} E_1 E_2 E_{IR}$, where $\chi^{(3)}$ is nonlinear susceptibility and $E_i$ is the amplitude of the optical field at 589, 569, and 2 338 nm, respectively, and subsequent UV generation.

The simultaneous detection of the forward-directed IR and UV radiation reveals that pronounced peaks and dips in the UV intensity curve follow the IR intensity peaks and drops with a certain delay, as shown in Fig. 5a. This lag can be partially explained by a technical reason, such as the PMT transit time. Also, an inertia of establishing the FWM fields might contribute to the observed temporal lag.

To quantitatively describe the observed similarity of temporal dependences, the Pearson coefficient is calculated [31]. The linear correlation between variables x and y can be characterized by the Pearson correlation coefficient r that is defined as

$$r = \frac{\sum_j (x_j - \langle x \rangle)(y_j - \langle y \rangle)}{\sqrt{\sum_j (x_j - \langle x \rangle)^2 \sum_j (y_j - \langle y \rangle)^2}},$$

where $\langle x_j \rangle$ and $\langle y_j \rangle$ are the means. Figure 5b shows the Pearson correlation coefficient r as a function of the time shift $\tau$ between two curves. This dependence has the form of damped oscillations. Initially, the coefficient r is small, but it grows with $\tau$ and reaches its maximum value ($r_{MAX}$ = 0.848) at $\tau_1$ = 32 ns. At larger time shift (32ns < $\tau$ < 76ns), the measured Pearson correlation coefficient $r$ monotonically drops to a perceptible negative value, $r_{MIN}$ = - 0.51 and then recovers to the secondary-maximum value.

Thus, the maximum measured Pearson correlation coefficient for the forward-directed IR emission at 2338 nm and UV radiation averaged over several realizations is $r_{MAX}$= 0.84±0.02. Correlation is not absolute, $r_{MAX}$<1, in part because dark noise contribution of both the UV and IR detectors, in part because of the nonlinear coupling between the applied laser and IR fields. Damping oscillations of the Pearson coefficient r are due to a certain periodicity of the spike occurrence that reflects temporal characteristics of the inversion, which is determined by excitation and relaxation processes.

Figure 6a demonstrates a certain similarity in appearance of chaotic peaks and dips in the temporal profiles of the forward-directed radiation at 330 and 2207 nm. The Pearson correlation coefficient r calculated for these profiles as a function of time shift $\tau$ between them is shown in Fig. 6b. This $r(\tau)$ dependence shows that there is a well-pronounced correlation between the temporal intensity features of the UV and the 2207 nm radiation at certain time shifts. However, its shape and the extreme r values differ from those shown in Fig. 5b for the optical fields at 330 and 2338 nm.

Firstly, as could be expected, the maximum Pearson correlation coefficient for the 330 and 2207 nm fields is smaller, the averaged value of the maximum correlation coefficient is $r_{MAX}$ = 0.61±0.02. Indeed, ASE at 2207 nm does not directly involved in the UV field generation. The appearance of intensity spikes at 2207 nm is determined by the inversion of population at the $4P_{3/2}$-$4S_{1/2}$ transition created by intensity spikes at 2238 nm. However, spontaneous decay at this transition introduces a certain uncertainty in appearance of intensity spikes at 2207 nm emission but not the UV field. This additional stochasticity of the 2207 nm lasing reduces the intensity correlation between the fields. Also, the complexity of the $r(\tau)$ dependence shown in Fig. 6b can be attributed to the influence of ASE at the cascade transitions. Since the dynamics of the 2207 nm field is determined by the temporal changes of the inversion at the $4P_{3/2}$-$4S_{1/2}$ transition, which depends on spontaneous and stimulated processes at the both $4D_{5/2}$-$4P_{3/2}$ and $4P_{3/2}$-$4S_{1/2}$ cascade transitions, the periodicity of the spike appearance is less pronounced than in the case of the 2338 nm lasing.

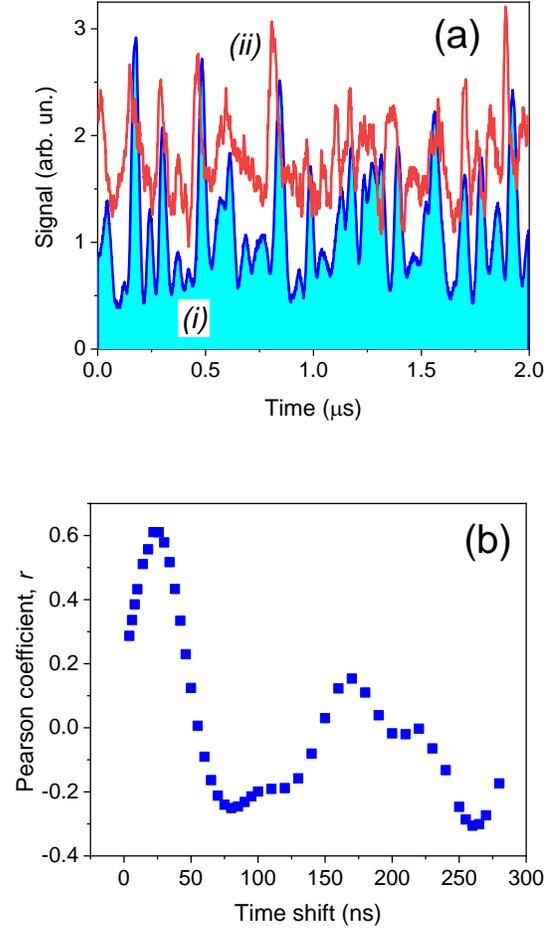

Fig. 6. Temporal profiles of (i) UV and (ii) IR emission at 2207 nm. The fixed-frequency pump lasers are tuned to the $3S_{1/2}(F=2)$ - $3P_{3/2}(F'=3)$ and $3P_{3/2}(F'=3)$ - $4D_{5/2}$ transitions, respectively. Curve (ii) is shifted vertically for a better visibility. (b) Correlation coefficient $r$ as a function of the time shift $\tau$ between the (i) and (ii) profiles.

Secondly, the observed intensity correlation between the 330 and 2207 nm fields also implies the presence of correlations between both IR fields. Considering that the Pearson coefficient r for the 330 and 2207 nm fields reaches its maximum value at the time shift of $\tau_2$ = 24 ns, we can conclude that the intensity peaks and dips at 2207 nm are delayed relative to the corresponding features of the 2338 nm field by $\Delta\tau = \tau_1 - \tau_2$ = 8 ns. This means that a certain time is needed for establishing directional radiation at 2207 nm after the inversion of population on the $4P_{3/2}$ - $4S_{1/2}$ transition is built by the

2338 nm field. We note that a delayed directional emission with respect to incoherent fluorescence was reported in [15].

Thus, the intensity spikes with durations shorter than the natural lifetimes of the 4P$_{3/2}$ energy level, as well as the time lag between the inversion creation at the 4P$_{3/2}$ - 4S$_{1/2}$ transition and the generation of directional radiation at 2207 nm, indicate a connection of the observed features of mirrorless lasing with the CEs, namely, with superfluorescence.

In conclusion, the directional IR and UV emission from sodium vapours excited to the 4D$_{5/2}$ level has been obtained and analysed.

Despite continuous-wave laser pumping, both the forward-directed mirrorless lasing at 2338 and 2207 nm generated on the cascade population-inverted 4D$_{5/2}$-4P$_{3/2}$ and 4P$_{3/2}$-4S$_{1/2}$ transitions as well as the UV radiation at 330 nm, which is produced by the FWM process, exhibit a spiking behaviour. The typical duration of isolated intensity spikes at 2338 nm is comparable, while at 2207 nm is noticeably shorter than the natural lifetime of the corresponding upper energy levels. This fact, as well as stochasticity of spikes and their certain delay, make it possible to associate the observed features with the collective effects.

The directional IR emission at 2338 nm and UV radiation reveal a high level of intensity correlation. Correlation also exists between cascade IR lasing at 2338 and 2207 nm as demonstrated indirectly via the UV field.

**Acknowledgments**. This work has been supported by the US Office of Naval Research Global under grant N62909-16-1-2113. We acknowledge fruitful discussions with Irina Novikova and Maria Chekhova. We thank the European Southern Observatory for the loan of a dye-laser system.